# Enumerating Stable Nanopores in Graphene and their Geometrical Properties Using the Combinatorics of Hexagonal Lattices

Sneha Thomas,[1,2] Kevin S. Silmore,[3] and Ananth Govind Rajan[2*]

[1]Department of Chemical Engineering, Indian Institute of Science Education and Research Bhopal, Bhauri, Madhya Pradesh 462066, India

[2]Department of Chemical Engineering, Indian Institute of Science, Bengaluru, Karnataka 560012, India

[3]Department of Chemical Engineering, Massachusetts Institute of Technology, Cambridge, Massachusetts 02139, United States

*Corresponding Author: Ananth Govind Rajan (E-mail: ananthgr@iisc.ac.in)

**ABSTRACT**

Nanopores in two-dimensional (2D) materials, including graphene, can be used for a variety of applications, such as gas separations, water desalination, and DNA sequencing. So far, however, all plausible isomeric shapes of graphene nanopores have not been enumerated. Instead, a probabilistic approach has been followed to predict nanopore shapes in 2D materials, due to the exponential increase in the number of nanopores as the size of the vacancy increases. For example, there are 12 possible isomers when $N$=6 atoms are removed, a number that *theoretically* increases to 11.7 million when $N$=20 atoms are removed from the graphene lattice. In this regard, the development of a smaller, exhaustive dataset of *stable* nanopore shapes can help future experimental and theoretical studies focused on using nanoporous 2D materials in various applications. In this work, we use the theory of 2D triangular "lattice animals" to create a library of all stable graphene nanopore shapes based on a modification of a well-known algorithm in the mathematical combinatorics of polyforms known as Redelmeier's algorithm. We show that there exists a correspondence between graphene nanopores and triangular polyforms (called polyiamonds) as well as hexagonal polyforms (called polyhexes). We develop the concept of a polyiamond ID to identify unique nanopore isomers. We also use concepts from polyiamond and polyhex geometry to eliminate unstable nanopores containing dangling atoms, bonds, and moieties. The exclusion of such unstable nanopores leads to a remarkable reduction in the possible nanopores from 11.7 million for $N$=20 to only 0.184 million nanopores, thereby indicating that the number of stable nanopores is almost two orders of magnitude lower and is much more tractable. Not only that, by extracting the polyhex outline, our algorithm allows searching for nanopores with dimensions and shape factors in a specified range, thus aiding the design of the geometrical properties of nanopores for specific applications. We also provide the coordinate files of the stable nanopores as a library to facilitate future theoretical studies of these nanopores.

**INTRODUCTION**

Nanoporous forms of two-dimensional (2D) materials, such as graphene, have recently received attention for several applications, such as gas separations,[1–5] ionic rejection,[6–10] nanofiltration,[11,12] and DNA sequencing.[13–15] Nanopores in graphene enable the selective permeation of molecules through the material, thereby allowing it to function as a membrane.[16,17] For this reason, nanopores in graphene have been studied extensively experimentally[18–23] and theoretically,[5,6,24–28] with recent advances involving high-resolution transmission electron microscopy (TEM) imaging[20,29] and multi-scale modeling of the etching of graphene nanopores.[30–32] As the size of the nanopore, quantified by the number of atoms etched from the lattice, increases, the number of possible shapes of nanopores that may exist increases exponentially.[32] Because of this challenge, researchers have not yet created a library of all possible nanopore shapes in graphene. Such a dataset could not only form a resource for future simulation studies of nanopores in graphene but could also enable automated identification of nanopores from TEM images, using machine learning. Accordingly, in this work, we develop a searchable library of stable nanopores in graphene, which allows for filtering nanopores using criteria such as the absence of dangling atoms/bonds/moieties, specified minimum/maximum values of the nanopore major and minor axes, and a given range for the nanopore's shape factor.

In the past, researchers have studied the formation of nanopores in graphene using ab initio density functional theory calculations,[33] reactive force field molecular dynamics simulations,[34,35] as well as kinetic Monte Carlo simulations.[30,32,36] The shapes and sizes of the nanopores (vacancy defects) formed play an important role in determining various properties of graphene, such as its magnetization,[37] gas separation efficacy,[28] and water desalination propensity.[8,9] This is because molecular and ionic rejection are highly sensitive to the topology of each nanopore. Apart from this, in applications like DNA sequencing, the shapes and sizes of nanopores dictate whether the presence of individual bases along the DNA can be precisely measured.[38] As an attempt to implement accurate control over the size of the nanopores, Zakharchenko and Balatsky studied the regrowth and healing of graphene nanopores by means of Monte Carlo simulations at various temperatures and pore size conditions.[39] Govind Rajan et al. cataloged the unique, most-probable shapes of nanopore isomers in graphene by combining density functional theory calculations, kinetic Monte Carlo simulations, and chemical graph theory.[32] Very recently, Sheshanarayana and Govind Rajan developed a machine learning framework based on nanopore structural features to predict the formation times and probabilities of arbitrary nanopore shapes in graphene.[40] Experimental studies have focused on unraveling the relative stability and performance of nanopores of different shapes in various materials.[41,42] The TEM-based investigations by Robertson et al.[20] and Huang et al.[29] have imaged many specific nanopores in graphene, providing a valuable dataset to infer the most-probable shapes of graphene nanopores. It follows that, for computational as well as experimental studies, understanding the plausible shapes of nanopores in graphene is a promising endeavor, from both fundamental and application-oriented standpoints.

Although several advances have been made in understanding and predicting nanopore topologies in graphene, as discussed above, the lack of a method to predict all possible stable nanopore shapes in a 2D lattice represents a knowledge gap in the literature. Addressing this gap to enable the screening of nanopore topologies is important due to the rapid increase in the possible number of nanopores as the number of atoms removed from the 2D lattice ($N$) increases. To this end, instead of using a probabilistic approach, we use concepts from the mathematical combinatorics of polyforms[43,44] for the enumeration of nanopores in graphene. Specifically, we employ a modification of an algorithm, initially proposed by Redelmeier,[44] to enumerate polyforms in rectangular lattices. This modification was introduced by Aleksandrowicz[45] and made Redelmeier's algorithm amenable to hexagonal lattices present in 2D materials, such as graphene. While it is known that the number of possible nanopore shapes increases exponentially with the number of carbon atoms removed ($N$) (e.g., from 12 possible isomers when $N$=6 to 11.7 million when $N$=20),[32] we show here that considering only *stable* nanopores (i.e., without dangling atoms, bonds, and moieties) leads to a significant reduction in the possible nanopore shapes. Furthermore, because the nanopore size is the most important criterion in membrane-based applications (e.g., molecular/ionic separations and DNA sequencing), we facilitate the searching of nanopores based on their size, by specifying a scheme to filter nanopores by major and minor axes lengths. Finally, because recent studies have highlighted the role of the nanopore shape in determining the performance of graphene membranes,[8,9] we also enable the searching of nanopores based on a shape factor. The rest of the paper is organized as follows. We first discuss the link between graphene nanopores and the combinatorics of lattice animals, such as polyiamonds and polyhexes. Next, we discuss the use of Redelmeier's algorithm that was originally developed for square lattices, in 2D materials possessing hexagonal lattices. Thereafter, we explain the use of symmetries to group the nanopore shapes, as well as methods to eliminate nanopores containing nonbonded atoms and dangling bonds/moieties. We end with a discussion on how the dataset may be searched for nanopores of various sizes and shapes.

**METHODS**

***The link between graphene nanopores and lattice combinatorics.*** Each nanopore in the graphene lattice can be considered as a "triangular lattice animal", i.e., a set of equilateral triangles sharing common sides with each other (see Figure 1), where each triangle represents a removed atom. Lattice animals on square lattices, also called polyominoes, have been extensively investigated in combinatorial and recreational mathematics.[46,47] On the other hand, two types of lattice animals exist on hexagonal lattices that are prominently featured in 2D materials, called polyiamonds and polyhexes. While a polyiamond is a polyform (a collection of polygons with common edges) consisting of equilateral triangles, a polyhex is a polyform consisting of regular hexagons. Polyhexes are frequently encountered in the study of polycyclic aromatic hydrocarbons,[48–50] while recently, the correspondence between polyiamonds and 2D nanopores[32] as well as colloidal clusters[51] has been alluded to. In Figure 1A, we depict a nanopore formed by removing 6 atoms from the graphene lattice. As seen in Figure 1B, a polyhex consisting of 7 hexagons captures the shape of this nanopore. Note that the center of each hexagon can be connected to form a polyiamond, as depicted in Figure 1C. Although it may not be apparent, polyiamonds form a convenient representation of a nanopore in graphene, because the center of each triangle in the polyiamond corresponds to an atom that was removed from the graphene lattice, as seen in Figure 1D. Thus, by representing the antimolecule of the nanopore as a polyiamond, there is a direct one-to-one correspondence between the various possible nanopore isomers in a graphene lattice and the polyiamonds that can exist on a hexagonal lattice. Indeed, the size of the nanopore, quantified as the number of carbon atoms etched, $N$, is the same as the number of triangles in the corresponding polyiamond.

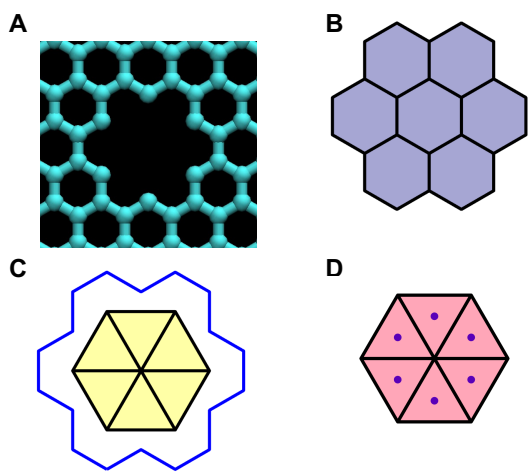

**Figure 1.** Correspondence between nanopores in graphene and lattice animals. (A) An example nanopore formed by removing $N=6$ atoms from the graphene lattice. (B) The polyhex shape corresponding to the nanopore shown in panel A. (C) The polyiamond shape corresponding to the nanopore shown in panel A, in yellow color. The blue colored outer boundary of polyhex is also shown for clarity. Note that the center of each hexagon in the polyhex corresponds to a vertex of the polyiamond. (D) The center of each triangle in the polyiamond corresponds to a removed carbon atom, as shown by the filled blue circles.

Based on the above observation, the different shapes of nanopores in a graphene sheet can be enumerated by finding all possible polyiamond shapes that can exist. For this purpose, one needs to make a distinction between *fixed* and *free* polyiamonds. *Fixed* polyiamonds are distinct if they cannot be transformed into each other via translations (Figure 2A). For example, the polyiamonds labeled as $a_1, a_2, \ldots, a_6$ in Figure 2A are distinct fixed polyiamonds, since they can be transformed into each other by rotation, but not by translation. On the other hand, *free* polyiamonds are distinct if they cannot be transformed into one another by a combination of reflections, rotations, and translations (Figure 2B). Thus, all the fixed polyiamonds labeled $a_1, a_2, \ldots, a_6$ in Figure 2A can be mapped to a single free polyiamond labeled as in Figure 2B. Hence, as seen in Figure 2, there are only 3 free polyiamonds for a nanopore of size $N=4$ (Figure 2B), but there are 14 fixed polyiamonds for the same value of $N$ (Figure 2A). Note that, due to the 12 symmetries present in graphene (6 rotational and 6 reflectional for space group *p6mm*), nanopores which are related by rotations, reflections, and translations, are equivalent to each other. Thus, in this work, we are interested in enumerating the *free* polyiamonds on a hexagonal lattice.

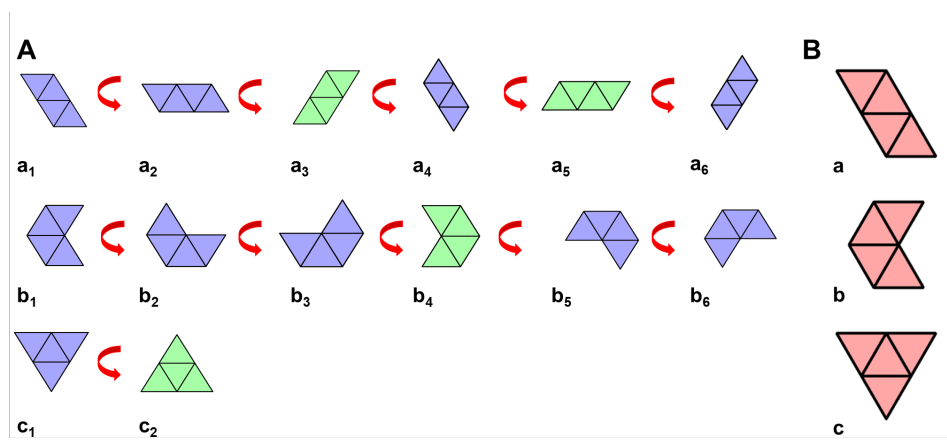

**Figure 2.** The distinction between fixed and free polyiamonds. (A) All 14 fixed polyiamonds of size $N=4$. Polyiamonds shown in green have their first triangle as an upright one and those shown in pink have their first triangle as an inverted one (see the section on the generation of free polyiamonds for more details). The red arrows indicate rotation in the counterclockwise direction by 60 degrees. (B) All 3 free polyiamonds of size $N=4$. Note that in panel A, there are a total of 14 fixed polyiamonds, where shapes $a_1$ to $a_6$, $b_1$ to $b_6$, and $c_1$ to $c_2$ are rotationally equivalent. Panel B shows the 3 free polyiamonds obtained by choosing one shape each from three groups of rotational isomers ($a_1$ to $a_6$, $b_1$ to $b_6$, and $c_1$ to $c_2$).

***Enumerating graphene nanopores without considering reflectional and rotational symmetries.***

Since the number of polyforms on a lattice increases exponentially with the size of the polyform, their enumeration presents a computational challenge.[44] In 1981, Redelmeier developed a rapid algorithm to generate all the *fixed* polyominoes on a square lattice.[44] (Recall that fixed polyominoes do not take into consideration reflectional and rotational symmetries.) Subsequently, Aleksandrowicz proposed a method to apply Redelmeier's algorithm to hexagonal lattices in order to generate fixed polyiamonds.[43] In this work, we have modified the polyEnumDraw code for fixed polyomino generation[52] to make it amenable for enumerating free polyiamonds. To this end, one starts with a single triangle at the origin, (0,0), and keeps on adding triangles till a polyiamond of the desired size is obtained. As shown in Figure 3A, the triangles are indexed by two numbers which denote the location of each triangle along the two principal Cartesian axes, labeled *a* and *b* here. Redelmeier's algorithm requires that no triangles below the initial triangle, as well as no triangle on the same row to its left, are considered (see the triangles with a dashed outline in Figure 3A). The transformation of the polyiamond depicted in blue color in Figure 3A to a polyhex is shown in Figure 3B, where the boundary of the corresponding polyhex is marked in blue color around the polyiamond. After removing the carbon atoms from the graphene lattice in the pattern specified by the polyiamond in panel B, the molecular structure of the corresponding nanopore is obtained (Figure 3C). Thus, we can rapidly obtain all the fixed polyiamonds corresponding to a given nanopore size. Nevertheless, two challenges remain with respect to enumerating nanopore shapes – firstly, we are interested in *free*, and not *fixed*, polyiamonds, and secondly, the polyiamonds generated using Redelmeier's algorithm contain "holes" that correspond to unphysical, nonbonded atoms in the graphene lattice, as well as lead to nanopores having dangling bonds and moieties. We next discuss the resolution of each of these challenges one by one.

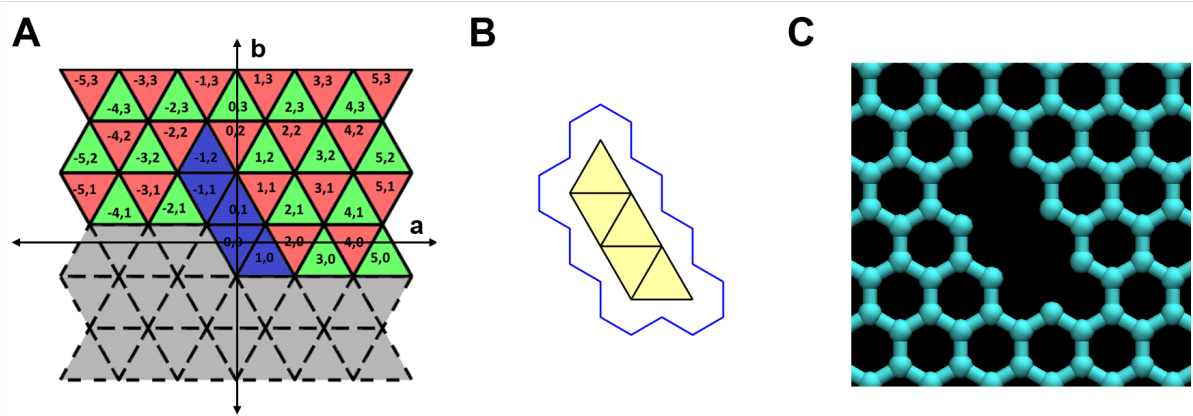

**Figure 3.** Illustration of Redelmeier's algorithm for generating fixed polyiamonds. (A) Hexagonal (triangular) grid for polyiamond generation using Redelmeier's algorithm. The triangle centers on the grid are labeled using the Cartesian coordinate system, where along axis '$b$', 1 unit = $\frac{1}{2\sqrt{3}s}$ and along axis '$a$', 1 unit = $\frac{s}{2}$, where $s$ is the side of the triangle. The algorithm requires that triangles below the initial triangle (i.e., the triangle at the coordinate (0,0), the origin), as well as triangles on the same row to its left (grey colored triangles with dashed outlines) are excluded. A sample polyiamond of size $N=5$ is generated, with its constituent triangles depicted in blue color. (B) The polyiamond generated in panel A is shown in yellow color along with the outer boundary of the corresponding nanopore rim, i.e., the polyhex shape corresponding to the nanopore generated. (C) Nanopore that is represented by the polyiamond shown in panels A and B in the graphene lattice.

***Removal of nanopores with nonbonded carbon atoms.*** As mentioned above, one needs to exclude polyiamonds with holes (Figure 4A), because the holes correspond to unphysical dangling atoms in the middle of a nanopore (Figure 4B). Indeed, any dangling, nonbonded atom in the lattice would get etched away rapidly due to the lack of bonds with other atoms.[32] Note that holes appear in polyiamonds only when $N > 9$.[53] To identify and eliminate nanopores with dangling atoms, we derived a mathematical relationship between the total number of vertices, $V$, the number of sixfold-shared vertices, $n_6$, and the number of triangles, $N$, in a polyiamond without holes, based on Euler's polyhedron formula:

$$N + 2 = V + n_6$$

The derivation of this equation is presented in the Supporting Information, Section S1. Any polyiamond that violates the above equation must necessarily contain a hole (i.e., a dangling carbon atom) and is removed from the list of possible fixed polyiamonds. This leads to a reduction in the number of nanopore isomers, as explained in the Results and Discussion section.

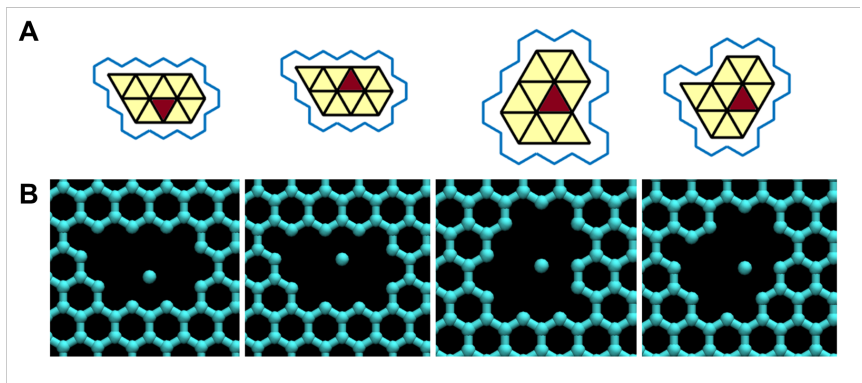

**Figure 4.** Free polyiamonds with holes. (A) The red-colored triangles in the polyiamonds are holes that got included in the shape while enumerating polyiamonds of size $N$=10. Note that there are 10 yellow-colored triangles. (B) By visualizing the corresponding nanopore shapes, the presence of nonbonded atoms is evident. These unremoved nonbonded atoms are represented by the red triangles in panel A.

***Identifying reflectional and rotational symmetries of the generated nanopores.*** As explained before, we are interested in *free* polyiamonds, since they correspond to unique nanopores in graphene. The total number of fixed polyiamonds is constituted by two types of polyiamonds: polyiamonds with an upright origin triangle (green-colored polyiamonds shown in Figure 2A) and polyiamonds with an inverted origin triangle (pink-colored polyiamonds shown in Figure 2A). As the enumeration of free polyiamonds from an initial set with a lower number of fixed polyiamonds would be more computationally efficient, rather than starting with the entire set of free polyiamonds, we explored the possibility of considering only fixed polyiamonds with an upright or inverted origin triangle. It is evident from Figure 2A that the fraction of polyiamonds with an upright origin triangle is lower (for $N$=5, 29%) than the fraction of polyiamonds with an inverted origin triangle. Although the fraction of polyiamonds with an upright origin triangle is lower, all free polyiamonds are not present inside this smaller set (see the Supporting Information, Section S2). Nevertheless, polyiamonds with an inverted origin triangle can be used for further reduction to free polyiamonds, as the map from fixed polyiamonds with inverted origin triangles onto free polyiamonds is surjective. We verified this by ensuring that the number of free polyiamonds predicted by our algorithm (starting with an inverted origin triangle) and that counted in previous works[53] match exactly. Accordingly, we started Redelmeier's algorithm exclusively with an inverted triangle at the origin (Figure 3A). Subsequently, we identified the reflectional and rotational symmetries in the generated fixed polyiamonds and chose canonical free polyiamonds by eliminating the nanopores related to them via any of the 12 symmetries present in graphene (six 60° rotations and six reflections). Thus, any polyiamond can have a maximum number of 12 symmetrical equivalents. Figure 5 depicts the symmetry operations that can be applied to a pentaiamond to obtain its rotational and reflectional symmetry-related fixed polyiamonds.

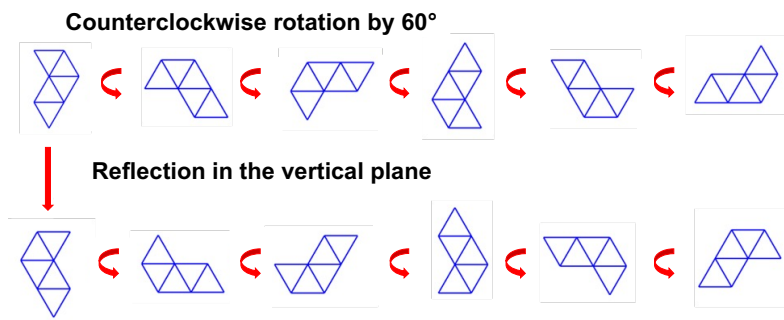

**Figure 5.** Rotational and reflectional symmetry operations to identify free polyiamonds from fixed ones. The pentaiamond shown here exhibits no internal symmetries (i.e., all 12 group actions yield a unique fixed polyiamond), meaning all twelve fixed polyiamonds must be identified and counted as one shape when enumerating free polyiamonds.

***Development of a unique ID for fixed polyiamonds to remove duplicate nanopores.*** Eliminating the symmetry-equivalent polyiamonds one-by-one via explicit comparison is computationally quite expensive. Therefore, we characterized *initially* each polyiamond by a "vertex repetition vector" (see below), so that they can be grouped and compared efficiently. In a polyiamond, a vertex can be shared by a maximum of six triangles. We define the vertex repetition vector of any polyiamond as $(n_1, n_2, ..., n_6)$, where $n_1$ denotes the number of vertices that are shared by only one triangle, $n_2$ denotes the number of vertices that are shared by two triangles, and so on. All the symmetrically equivalent nanopore shapes (i.e., fixed polyiamonds that map to the same free polyiamond) necessarily have the same vertex repetition vector, as rotational and reflectional operations do not affect vertex topology. Thus, the fixed polyiamonds were grouped based on their vertex repetition vector and only afterward compared based on symmetry operations to reduce the set of the fixed polyiamonds to that of the free polyiamonds. *Subsequently*, rather than applying symmetry operations to the coordinates of the polyiamond vertices and making comparisons, we developed a unique ID for each fixed polyiamond, which under specific operations captures the symmetry operations of rotation and reflection that collapse all symmetry-equivalent fixed polyiamonds into a single free polyiamond. The unique ID was the sequence of numbers representing the number of times each rim vertex is shared, in the consecutive counterclockwise order of the rim vertices, starting with the leftmost vertex of the bottommost row of the triangles. As a result, all the rotation isomers of a particular fixed polyiamond can be obtained by cyclic permutations of the ID assigned to that polyiamond. In other words, all rotational isomers of a particular fixed polyiamond can simply be detected by determining whether the ID of the candidate rotational isomer and the given polyiamond are cyclic permutations of each other (see Figure 6A-E). Similarly, all reflectional isomers of a particular fixed polyiamond with an assigned ID can be detected by determining whether the *reversed* ID of the candidate rotational isomer and the original ID of the given polyiamond are cyclic permutations of each other (see Figure 6F-I). For this method to work, it is important to list all the rim vertices in their order of connectivity. The algorithm used to obtain the order of connectivity is discussed in detail in the Supporting Information, Section S3. To increase the efficiency of the code and speed up the searching of the unique nanopore shapes, the comparison to detect unique free polyiamonds was only done within the groups created as per the vertex repetition vector, that was defined earlier. In total, the above procedure reduced efficiently the whole set of fixed polyiamonds without holes initially generated (as discussed in the last section) to free polyiamonds without holes.

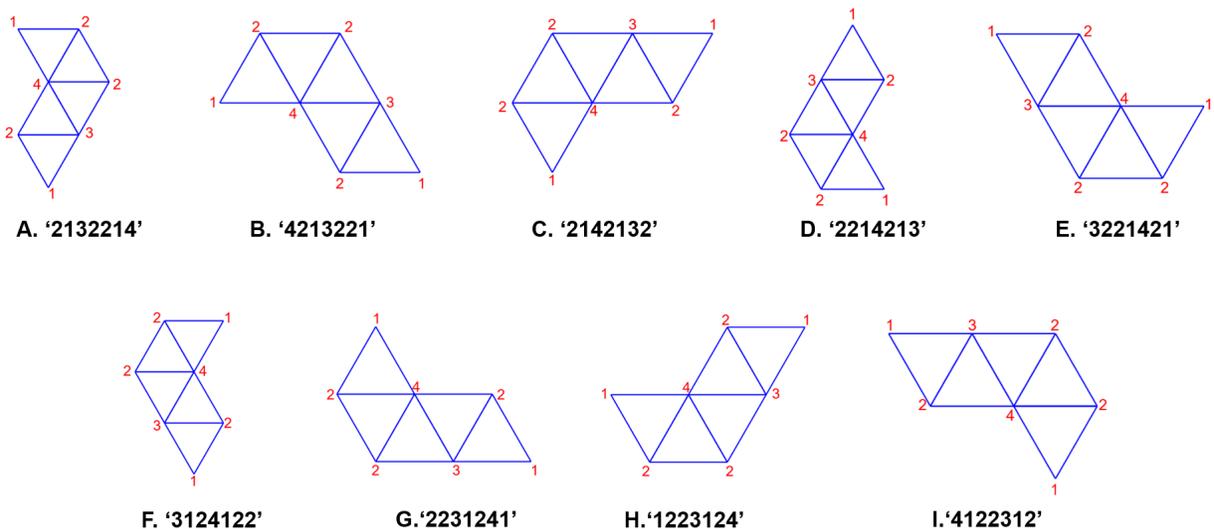

**Figure 6.** Assignment of IDs to fixed polyiamonds starting with an inverted triangle, for the purpose of mapping fixed polyiamonds to free ones. Each ID is generated using the sequence of numbers representing the number of times each rim vertex is shared, in the consecutive counterclockwise order of the rim vertices, starting with the leftmost vertex of the bottommost row of the triangles. Note that the IDs of the fixed polyiamonds (rotational isomers) in panels A through E and F through I are cyclic permutations of each other, respectively, and thus they all correspond to the same free polyiamond. The IDs of the fixed polyiamonds in panels F through I (reflectional isomers) are reversed cyclic permutations of the IDs for panel A through E.

***Elimination of graphene nanopores with singly bonded carbon atoms and dangling moieties.***

Carbon atoms in the graphene lattice have coordination numbers ranging from one to three (see, e.g., the nanopore depicted in Figure 7A). Previously, we eliminated carbon atoms with zero coordination numbers (i.e., nonbonded carbon atoms) from the list of nanopores when we excluded all polyiamonds with holes. However, a "dangling bond", i.e., a carbon atom that is singly bonded is also unstable and will be readily etched during nanopore formation.[32]

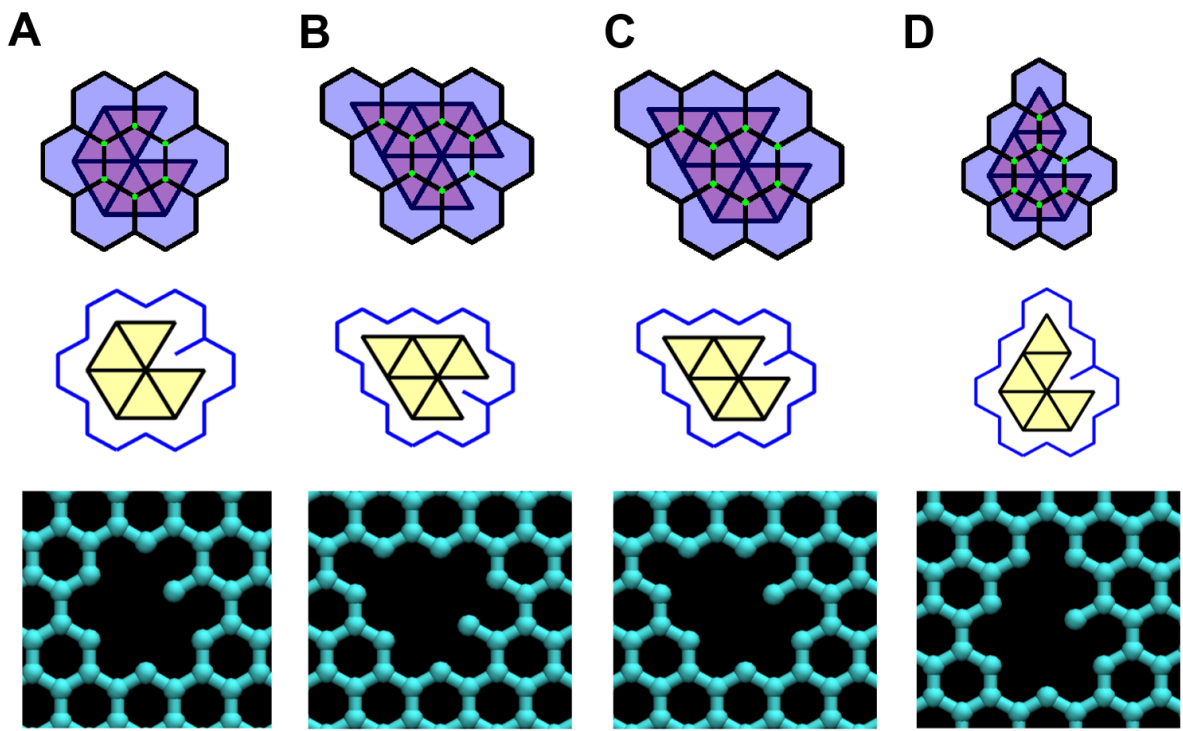

**Figure 7.** Graphene nanopores with singly bonded carbon atoms (dangling bonds). Polyhexes and polyiamonds (top), polyhex rims and polyiamonds (middle), and atomic structures (bottom) of nanopores of size $N=5$ (A) and $N=6$ (B-D) having dangling bonds. The green-colored points denote the points inside the polyhex (these points are hexagonal vertices that are trivalent). In each case, a vertex shared by 5 triangles is present. Note that out of the 4 free polyiamonds of size $N=5$, only one polyiamond represents a nanopore with a singly bonded carbon atom (dangling bond), as shown in panel A. Similarly, out of the 12 free polyiamonds of size $N=6$, only three polyiamonds correspond to a nanopore with a singly bonded carbon atom (dangling bond), as shown in panels B-D.

Thus, we developed a methodology to screen and remove nanopores with singly bonded carbon atoms, by making use of polyiamond connectivity and polyhex shapes. As is visually clear in Figure 7, one way to eliminate nanopores with a singly bonded carbon atom (dangling bond) is to find polyiamonds containing vertices that are shared by 5 triangles. Using this procedure, polyhexes without dangling bonds were isolated and stored as a library for each *N*. Exclusion of nanopores with dangling bonds reduced the dataset size enormously and made it more manageable, as explained in the Results and Discussion section. Besides nanopores containing nonbonded and singly bonded carbon atoms, starting from *N*=13, there were other highly improbable pores present in the dataset, featuring dangling moieties of various shapes, as shown in Figure 8. These nanopores contained one (Figure 8A), two (Figure 8B), or more dangling hexagons inside the rim. We eliminated such nanopores by deleting the polyhexes with self-intersecting boundaries (i.e., where the same edge is repeated twice; see the Supporting Information, Section S4), as edges overlap on the boundary if and only if there is a dangling moiety. Finally, the boundary of each polyhex was determined by arranging the vertices of the polyhex based on the required angle and distance between consecutive vertices in a polyhex. The order of connectivity of the rim atoms is required to calculate different properties to characterize the nanopore. The algorithm to obtain the order of connectivity of the polyhex rim is mentioned in detail in Section S4 of the Supporting Information.

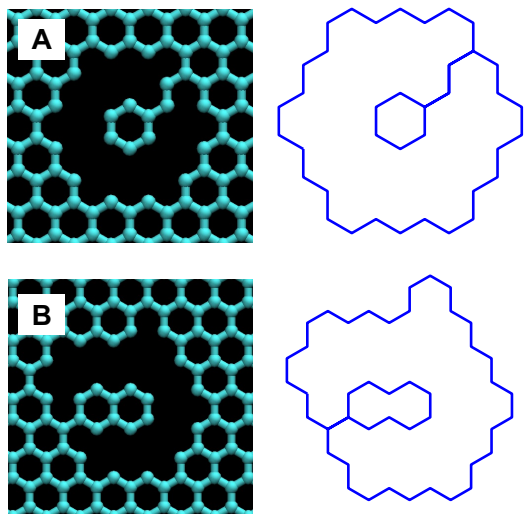

**Figure 8.** Examples indicating the presence of dangling moieties in nanopore shapes in graphene. (A) A nanopore of size *N*=16 with a dangling moiety. (B) A nanopore of size *N*=18 with a dangling moiety. In both A and B, the blue lines mark the rim of the nanopore.

**RESULTS AND DISCUSSION**

***Number of stable nanopores in graphene.*** Using the methods described above, we generated a library of stable nanopores in graphene for sizes ranging from $N=1$ to $N=20$. The structures of these nanopores and the code used to generate them are available in an open-source code repository (see the Code Availability statement). Due to the systematic elimination of unphysical nanopores structures (e.g., those containing nonbonded atoms and dangling bonds/moieties), we were able to considerably reduce the size of the nanopore dataset. Figure 9 depicts the workflow used in this work to generate all the stable nanopores in graphene for a given number of atoms removed and the corresponding number of shapes obtained for an example size of $N=15$. As shown in Figure 9, when $N = 15$, the number of fixed polyiamonds is 886,160 and is reduced to 811,346, when fixed polyiamonds with holes are eliminated. The consideration of nanopore symmetries led to a more than tenfold reduction in the number of shapes to about 67,716. Finally, the elimination of nanopores with dangling bonds resulted in another order of magnitude decrease in the number of physically possible nanopores to only 4972, a number which further reduced to 4649 upon removing nanopores with dangling moieties. Thus, our physical understanding-based mathematical approach has the potential to make tractable the exponentially increasing number of nanopore shapes (free polyiamonds) encountered in graphene by considering only stable nanopores. Furthermore, Figure 10A depicts the total number of nanopore shapes and the number of stable nanopore shapes for $N$ ranging from 1 through 20. In addition, Figure 10B shows a plot of the fraction of the nanopores that are stable for each value of $N$ considered. One can see clearly that the fraction of stable nanopores is as low as 0.016 for $N=20$ and is continually decreasing as the size of the nanopore increases. Physically, the combinatorial increase in the number of positions for dangling bonds and dangling moieties leads to this tremendous decrease in the fraction of stable nanopores as *N* increases. This finding provides confidence that our approach can make the study of graphene nanopores tractable by making it possible to consider only the much smaller subset of stable nanopores, rather than considering the massive set of all free polyiamond shapes.

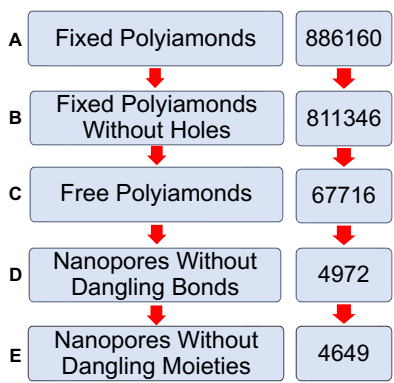

**Figure 9.** Workflow to generate nanopore shapes in graphene. The number of shapes found at each step is also indicated for the example case of *N*=15. (A) Generation of fixed polyiamonds. (B) Removal of polyiamonds with holes, i.e., nanopores with nonbonded atoms. (C) Conversion of fixed polyiamonds to free ones using symmetry considerations. (D) Exclusion of nanopores with dangling bonds. (E) Exclusion of nanopores with dangling moieties.

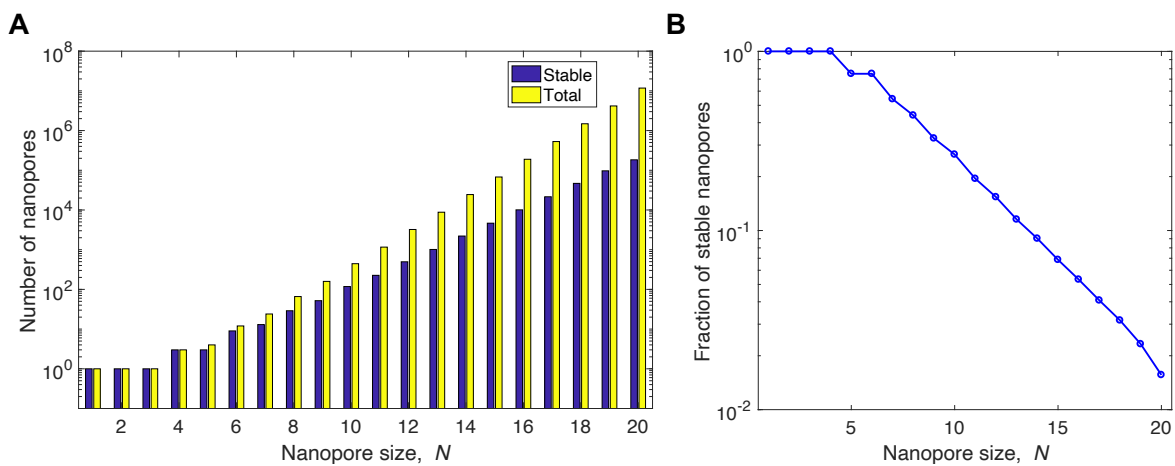

**Figure 10.** Plot of the total number of nanopores and the nanopores without any dangling bonds and moieties, as a function of *N*. (A) The navy bars indicate the number of stable nanopores, i.e., nanopores without dangling bonds and dangling moieties. The yellow bars indicate the total number of nanopores, i.e., all free polyiamonds without holes, which includes nanopores containing dangling bonds and dangling moieties. (B) Plot of the fraction of stable nanopores as a function of *N*. Both plots use a logarithmic scale for the vertical axis.

***Distribution and filtering of the structural features of stable nanopores.*** Computational screening of nanopores should prove to be a vital tool for theorists in predicting structure-property relationships of nanoporous media for a variety of applications. Recently, Bondaz et al. addressed the limitations in screening and understanding the transport characteristics of nanopores (caused by the expensive computational cost imposed by molecular simulations) using rapid screening methods based on pore shape and size, as characterized by the pore limiting diameter (PLD),[54] i.e., the diameter of the largest circle that can fit inside a nanopore. On the other hand, experimentalists may want to examine the atomic structures of nanopores like what they may observe in microscopy studies. To aid such chemical discovery efforts, we developed a search function based on specified ranges of the: (i) major and minor axes lengths of a nanopore and (ii) shape factor defined as the ratio of the equivalent circular diameter and major axis length of a nanopore. The equivalent circular diameter, $d_{circ}$, is simply the diameter of the circle having the same area as the nanopore shape, with the latter determined by the number of incomplete hexagons in the graphene lattice (i.e., the area of the corresponding polyhex). The major axis length, $d_{max}$, is the length of the longest line that can be drawn through an object. The major axis endpoints were found by testing every pair of points along the boundary of the nanopore. The shape factor, $S$, is defined as:

$$S = \frac{d_{circ}}{d_{max}}$$

Further, the minor axis was found by determining the shortest rim-to-rim distance that is perpendicular to the major axis of the polyhex shape. In Figure 11, we depict the nanopores with maximum and minimum shape factors, major axis length, and minor axis length. As can be seen therein, expectedly, the pores with the longest major axis are elongated pores (panels $A_2$, $B_1$, and $C_1$), and they also typically have the shortest minor axis and a low shape factor (0.4 to 0.6). In the table presented in Figure 11D, the nanopores having maximum major axis length and minimum shape factor are indeed the same. However, such pores have a low probability to form, as shown recently in the work by Sheshanarayana and Govind Rajan.[40] On the hand, pores with the longest minor axis (panels $A_3$, $B_3$, and $C_2$) are more circular and have shape factors between 0.7 and 0.8. This tendency of pores with longer minor axis to be circular should reflect in them also having the maximum shape factors, but, due to the possibility of curved (Figure 11$C_2$) and horseshoe-shaped (Figure 11$B_2$ and 11$C_4$) nanopores as $N$ increases, this relationship is not very much visible among the maximum and minimum values of shape factors given in the table presented in Figure 11D.

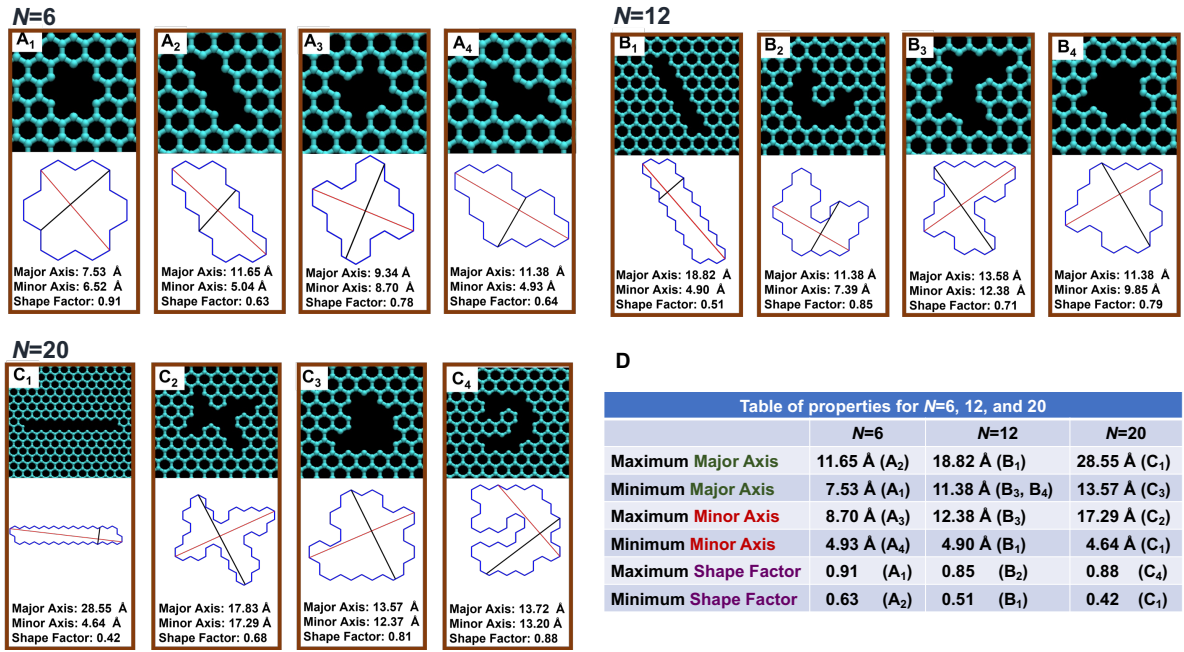

**Figure 11.** Searching through the nanopore database using geometric properties such as the major axis length, minor axis length, and shape factor. ($A_1$-$A_4$) Nanopores for $N$=6 with maximum and minimum values of properties. ($B_1$-$B_4$) Nanopores for $N$=12 with maximum and minimum values of properties. ($C_1$-$C_4$) Nanopores for $N$=20 with maximum and minimum values of properties. In each case, the nanopore outline is shown in blue, the major axis in green, and the minor axis in red. (D) A table summarizing the minimum and maximum property values and nanopore corresponding to each value.

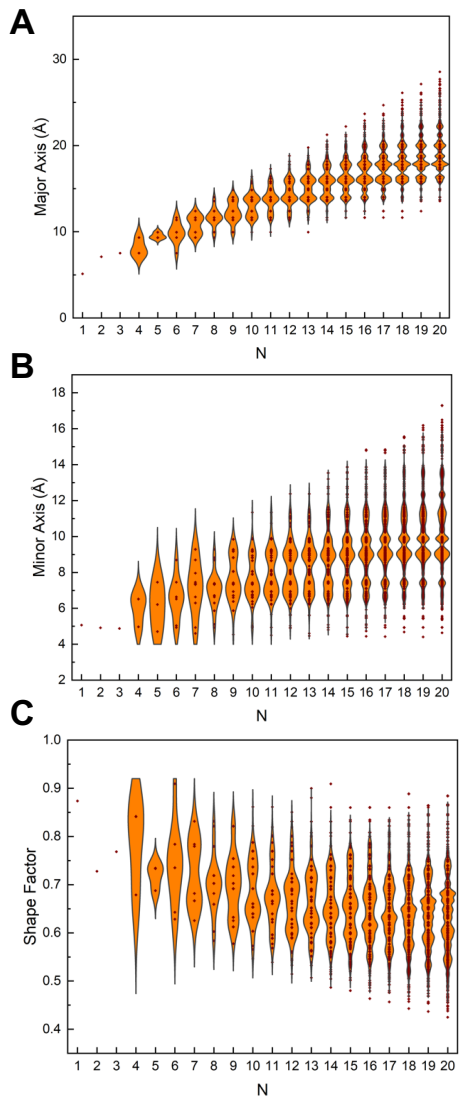

**Figure 12.** Violin plots of the geometrical properties of graphene nanopores (major axis length, minor axis length, and shape factor) with at most 20 atoms removed. The vertical extent of the violin plot shows the range of the data, while the width of the violin plot denotes the frequency of the datapoints at each value. (A) Violin plot of the major axes lengths. (B) Violin plot of the minor axes lengths. (C) Violin plot of the shape factors.

To obtain further insight into the trends exhibited by the major axes, minor axes, and shape factors of graphene nanopores, we present in Figure 12 the density of values of properties with varying $N$. We note that the minimum length of the minor axis for different $N$ are almost the same (Figure 12 B) due to the similarities among the maximally elongated shapes (Figure 11$A_2$, $B_1$, and $C_1$). On the other hand, the minimum and maximum major axes lengths, as well as the maximum length of the minor axis, increase with $N$, as the size of the pore increases (Figure 12A and B). Finally, the shape factor values range from 0.4 to 0.9, with smaller nanopores having a more uniform distribution of shape factors, and larger nanopores exhibiting more instances of lower shape factors, i.e., elongated nanopores. Overall, the minor axis lengths for nanopores with at most 20 atoms removed range from 5-17 Å and the major axis lengths for the same nanopores range from 5-28 Å.

Before concluding, we note the advantages of the nanopore searching feature made available here in the context of nanopore design studies using machine learning.[9,40,54–56] As mentioned before, Bondaz et al. used the concept of PLD as it can capture the protrusions and irregularities of the pore shape.[54] In a similar vein, the properties that we have enlisted can help the screening process of finding suitable pores, albeit in a faster manner. Since PLDs can be seen as a function of the major and minor axes lengths as well as the shape factor, after screening the nanopores using such readily calculated geometric shape properties, the PLDs of the screened pores can be calculated using more expensive methods of optimization and/or image processing. Such a two-step procedure could be very helpful while handling the massive data of nanopores which is inevitable for studies that will seek to use reinforcement or machine learning strategies for the inverse-design of 2D nanopores.

**CONCLUSIONS**

In this work, we developed algorithms to generate and characterize stable 2D nanopore shapes corresponding to a fixed number of atoms removed from a hexagonal lattice. We used an algorithm from combinatorial mathematics due to Redelmeier to generate the various nanopore shapes. We introduced concepts such as fixed and free polyiamonds and discussed how the latter relates to nanopores in graphene. We showed that elimination of nanopores with nonbonded atoms and dangling bonds/moieties can lead to a drastic reduction in the number of shapes, thereby making the dataset tractable for future studies. Finally, we implemented a search method for nanopores to obtain nanopores with sizes quantified by their major/minor axis length and their shapes quantified by their shape factor, to aid future studies. Our study will find use in the computational discovery of nanopore shapes aimed at a specified target application, such as DNA sequencing, gas separation, or water desalination. Furthermore, in the future, our nanopore library can be used to rapidly identify nanopore shapes from TEM imaging using machine-learning based tools. We hope that our study motivates the investigation of the effect of the shape and size of nanopores in 2D materials on various physicochemical properties of interest.

## SUPPORTING INFORMATION

Proof of the theorem used to detect and eliminate polyiamonds with holes, proof by contradiction that all the free polyiamonds are not necessarily present inside the polyiamond set generated by Redelmeier's algorithm starting from (0,0) with an upright triangle, algorithms used to obtain the order of connectivity of the vertices in a polyiamond and to assign an ID to each polyiamond, and algorithm used to obtain the order of connectivity of vertices in a polyhex.

## CODE AVAILABILITY

The MATLAB code used to generate the nanopore library, the MATLAB code to search the nanopore database, and a library of XYZ files of all nanopores without dangling atoms, bonds, and moieties from $N$=1 through $N$=20 are available at https://github.com/agrgroup/StableNanopores.

## ACKNOWLEDGEMENTS

A.G.R. acknowledges financial support from the National Supercomputing Mission, which is coordinated by the Department of Science and Technology (DST) and the Department of Electronics and Information Technology (DeitY). The authors thank the Supercomputer Education and Research Centre at the Indian Institute of Science for computational facilities.

**For Table of Contents Use Only**

*Enumerating Stable Nanopores in Graphene and their Geometrical Properties Using the Combinatorics of Hexagonal Lattices*

*Sneha Thomas, Kevin S. Silmore, and Ananth Govind Rajan*

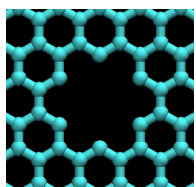
**2D nanopores**

**Polyiamond**
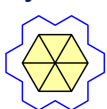

**Polyhex**
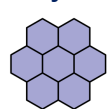

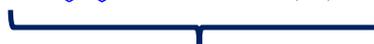
Unique & stable shapes
Geometrical properties

# Supporting Information for:

# *Enumerating Stable Nanopores in Graphene and their Geometrical Properties Using the Combinatorics of Hexagonal Lattices*

*Sneha Thomas,[1,2] Kevin S. Silmore,[3] and Ananth Govind Rajan[2]\**

[1]Department of Chemical Engineering, Indian Institute of Science Education and Research Bhopal, Bhauri, Madhya Pradesh 462066, India

[2]Department of Chemical Engineering, Indian Institute of Science, Bengaluru, Karnataka 560012, India

[3]Department of Chemical Engineering, Massachusetts Institute of Technology, Cambridge, Massachusetts 02139, United States

**\*Corresponding Author:** Ananth Govind Rajan **(E-mail:** ananthgr@iisc.ac.in)

## Table of Contents

**S1. Proof of the theorem used to detect and eliminate polyiamonds with holes**

**Theorem:** For any polyiamond without holes, two added to the number of triangles in the polyiamond (i.e., the size of nanopore, $N$) is the sum of total number of vertices, $V$ and the number of vertices getting shared by six triangles (i.e., 6-valent interior vertices), $n_6$.

$$N + 2 = V + n_6$$

**Proof:** Each vertex in a polyiamond can be assigned a degree or valency, which is defined as the number of connected edges coming out of it. The least degree or valency of a vertex in a polyiamond is two and the greatest degree possible is six, as demonstrated in Figure S1 using an example polyiamond. It can also be seen in Figure S1 that bivalent vertices are not shared between triangles, whereas trivalent vertices are shared between two triangles, tetravalent vertices are shared between three triangles, and so on.

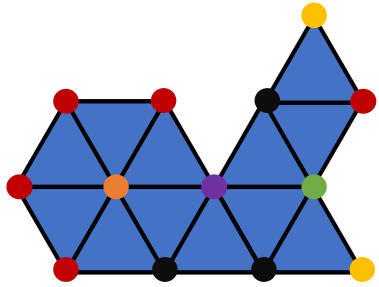

**Figure S1.** Vertices with different valencies, i.e., number of connected edges, are marked with different colors: vertices shared by 1 triangle, i.e., bivalent vertices (yellow colored); vertices shared by 2 triangles, i.e., trivalent vertices (red colored); vertices shared by 3 triangles, i.e., tetravalent vertices (black colored); vertices shared by 4 triangles, i.e., pentavalent vertices (green colored); vertices shared by 5 triangles, i.e., hexavalent exterior vertices (violet colored); and vertices shared by 6 triangles, i.e., hexavalent interior vertices (orange colored).

The total number of vertices in a polyiamond is the count of all these types of vertices. Let the number of vertices shared by 1 triangle (bivalent vertices) be $n_1$, that by 2 triangles (trivalent vertices) be $n_2$, that by 3 triangles (tetravalent vertices) be $n_3$, that by 4 triangles (pentavalent vertices) be $n_4$, that by 5 triangles (hexavalent exterior vertices) be $n_5$, and that by 6 triangles (hexavalent interior vertices) be $n_6$. Then, the total number of vertices counted uniquely, $V$, is given as:

$$V = n_1 + n_2 + n_3 + n_4 + n_5 + n_6 \quad (1)$$

To relate $n_1$ through $n_6$ with $N$, we consider a polyiamond, and separate out each triangle in it, as shown via an example in Figure S2A-B.

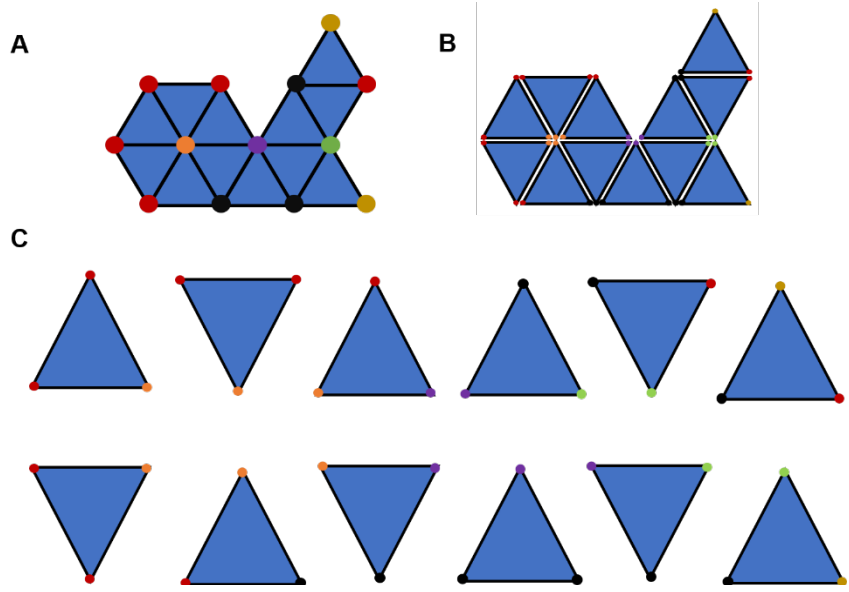

**Figure S2.** (A) Sample polyiamond ($N=12$) with marked vertices. (B) The total number of vertices, with repeats, is three times 12, i.e., 36. (C) The total number of vertices, with repeats, can also be obtained by segregating the triangles in the polyiamond and taking the valency-weighted sum of the number of each type of vertex.

Since each triangle has 3 vertices, the separated-out triangles will have a total of $3N$ vertices. Further, note that $n_j$ vertices are shared by $j$ triangles, such that after separating out the triangles, we will obtain $jn_j$ vertices of valency $j$, as shown in Figure S2C. It follows that:

$$3N = 1n_1 + 2n_2 + 3n_3 + 4n_4 + 5n_5 + 6n_6 \qquad (2)$$

Now, polyiamonds can be considered as planar graphs, so that Euler's polyhedral formula can be applied. Euler's polyhedral formula states that:[1]

$$V - E + F = 2 \qquad (3)$$

where $V$ is the number of vertices, $E$ the number of edges, and $F$ the number of faces or regions in the graph. Note that Euler's formula is only applicable to polyiamonds without holes.

The number of edges can be related to the number of vertices.[2] As mentioned earlier, each vertex has a minimum of 2 edges and a maximum of 6 edges connected to it, which also correspond to their degrees or valencies. Hence, a bivalent vertex will have 2 edges coming out of it, trivalent vertices will have 3 vertices coming out of it and so on. In general, with respect to $j$-valent vertices, $j$ times the number of $j$-valent vertices will be the total number of edges coming out of all the $j$-valent vertices. Considering all types of vertices with valencies upto 6 in a polyiamond, we have:

$$E = 0.5(2n_1 + 3n_2 + 4n_3 + 5n_4 + 6n_5 + 6n_6) \qquad (4)$$

where, since each edge is shared by 2 vertices, the coefficient of 0.5 appears in Equation (4). Next, consider that a polyiamond has $N+1$ regions, consisting of the $N$ triangles and the region outside the polyiamond, so that:

$$F = N + 1 \tag{5}$$

Substituting Eq. (1) for $V$, Eq. (4) for $E$, and Eq. (5) for $F$ in Euler's formula and simplifying, we obtain:

$$2N - 2 = n_2 + 2n_3 + 3n_4 + 4n_5 + 4n_6 \tag{6}$$

Subtracting Eq. (2) from Eq. (6), we get:

$$N + 2 = n_1 + n_2 + n_3 + n_4 + n_5 + 2n_6 \tag{7}$$

Using Eq. (1) in Eq. (7), we obtain the result

$$N + 2 = V + n_6 \tag{8}$$

which is the desired theorem. Note that this equation is not valid for polyiamonds with holes. The condition represented by Eq. (8) for polyiamonds without holes is easy to apply in our code because $N$ and $V$ are easily known, and $n_6$ is pre-determined for generating the unique ID of a polyiamond, as explained in the paper.

**S2. Proof by contradiction that all the free polyiamonds are not necessarily present inside the polyiamond set generated by Redelmeier's algorithm starting from (0,0) with an upright triangle**

Figures S3A and S3B show all the fixed polyiamonds corresponding to the same free polyiamond. The fixed polyiamonds in Figure S3A were generated starting at an inverted triangle and the fixed polyiamonds in Figure S3B were generated starting at an upright triangle. As can be inferred, the percentage of polyiamonds starting with upright triangles (Figure S3B) is less than or equal to the percentage of population of polyiamonds starting with inverted triangles (Figure S3A). Thus, it would have been computationally efficient to consider only the polyiamonds starting with upright triangles to determine the free polyiamonds. However, as shown in Figure S3C, there can be polyiamonds which cannot be generated starting with an upright triangle. Therefore, in this work, we used the set of fixed polyiamonds starting with inverted triangles to obtain the set of free polyiamonds.

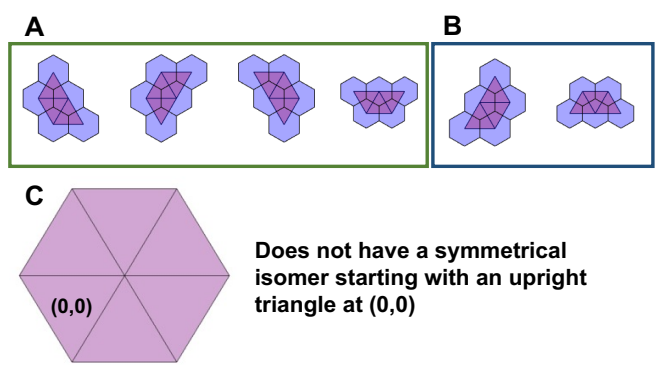

**Figure S3.** (A-B) All the fixed polyiamonds starting with an inverted triangle (A) and an upright triangle (B) corresponding to the same example free polyiamond. (C) An example

polyiamond generated starting with an inverted triangle that cannot be obtained starting with an upright triangle.

**S3. Algorithms used to obtain the order of connectivity of the vertices in a polyiamond and to assign an ID to each polyiamond**

One needs to determine the order of connectivity of the vertices in a polyiamond to assign an ID to each nanopore. To this end, we first find the boundary edges, which is done by traversing all the edges of the polyiamond and removing the edges which occur twice (because all edges except the boundary edges are shared by two triangles). Each edge is indexed as $(V_1, V_2)$, where $V_1$ and $V_2$ are two arbitrary vertices. Note that $(V_1, V_2)$ is the same edge as $(V_2, V_1)$. We now have a list of the boundary edges. We start with one of them and successively search the rest of the edges to find a new edge that is connected to the previous one via a shared vertex. We repeat until the boundary is closed and thus obtain a particular ordering of the vertices in the polyiamond.

We now illustrate the algorithm used to assign an ID to each polyiamond, using an example polyiamond of size $N=3$. As shown in Figure S4, the list of unique vertices is determined, and the vertices are numbered (Figure S4A). Subsequently, the vertex repetition number of each vertex is determined (Figure S4B) and the ordering of the vertices in the polyiamond is found out using the algorithm described above. Subsequently, the vertex repetition numbers are arranged in the order of the vertices to obtain the ID (Figure S4C). As discussed in the paper, the cyclical symmetry of the ID of a given fixed polyiamond with the IDs of other symmetrical fixed polyiamonds is used to reduce the whole set to free polyiamonds.

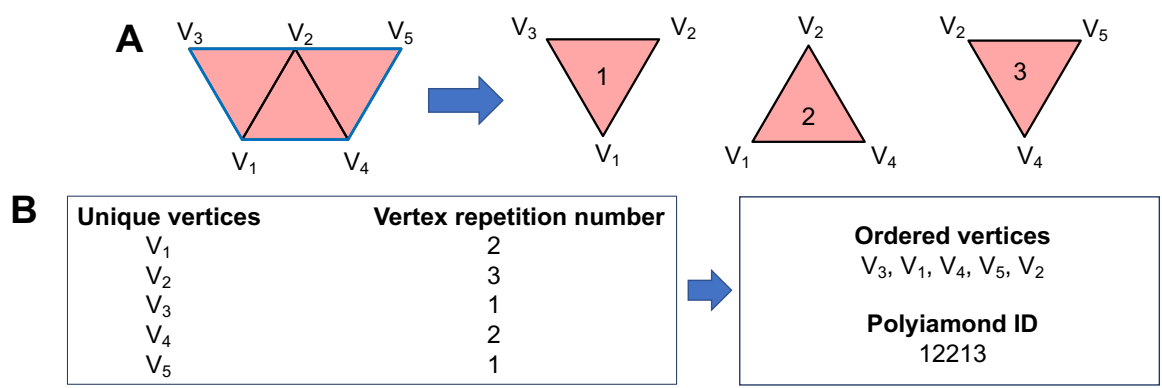

**Figure S4.** Demonstration of the algorithm used to assign an ID to a polyiamond using an example polyiamond of size $N=3$. (A-B) The list of unique vertices is obtained and the number of times each vertex is repeated is determined. (B) The vertices are ordered in the consecutive counterclockwise order of the rim vertices, starting with the leftmost vertex of the bottommost row of the triangles and the vertex repetition numbers are arranged in this order to obtain the ID of the polyiamond.

**S4. Algorithm used to obtain the order of connectivity of vertices in a polyhex**

The algorithm used here is almost same as the one discussed above for polyiamonds. Note that, as seen in Figure S5, due to combinatorial probabilities, physically improbable shapes with dangling moieties can also be present in the database. Although such nanopores are later removed from the nanopore library, if needed, to calculate their properties without any exemptions, it is important to find the boundary or rim of the polyhex of such complex or exceptional cases also.

If there are dangling moieties in a nanopore (Figure S5A-C), traversing through at least one polyhex edge twice will be necessary (Figure S5D). Hence, in the case of polyhexes, ($V_1$, $V_2$) is not considered the same edge as ($V_2$, $V_1$). To retain both edges in the list, after finding the boundary edges, the methodology used to eliminate all the *inner* hexagonal vertices, i.e., those that do not lie at the boundary, is different. Thus, rather than removing all edges which occur twice, edges containing all hexagonal vertices which occur thrice, or which are shared by 3 hexagons are removed.

The other difference is that, to tackle exceptional cases as shown in Figure S5A-C, at vertices where there are 3 edges coming out of it (vertex marked as 37), while traversing, priority is given to traverse the twice-occurring edge first. Also, since ($V_1$,$V_2$) and ($V_2$,$V_1$) are considered different, we avoid ($V_2$,$V_1$) getting traversed right after ($V_1$,$V_2$) by checking and keeping the last-added vertex and the newly added vertex always different. The rest of the algorithm is the same as that for polyiamonds. We start with any one of the edges and successively loop through the rest of the edges until we find a common vertex between the two edges. The process is then repeated until the boundary is closed.

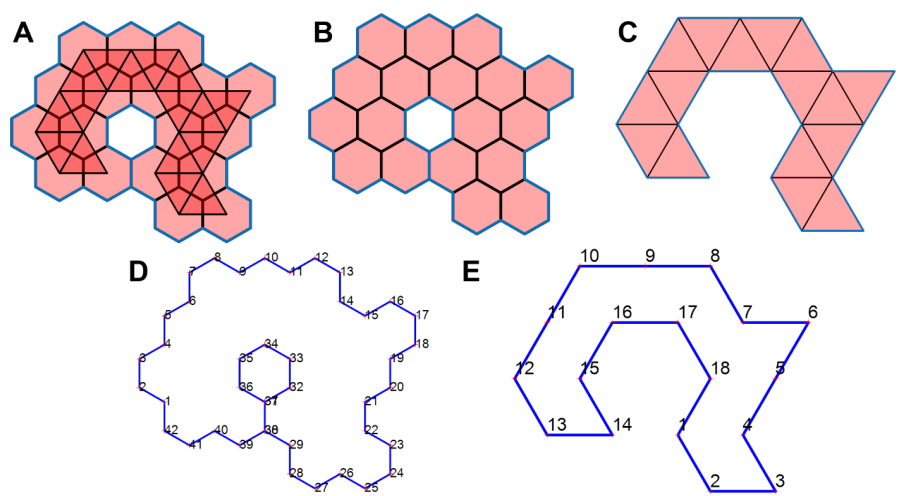

**Figure S5.** (A-C) Polyhex and polyiamond (A), polyhex only (B), and polyiamond only (C) corresponding to an example nanopore with a dangling hexagonal moiety. The boundary edges which are not shared with between multiple hexagons are marked in blue. The inner edges which are shared by two hexagons are marked in black. Edges are similarly marked for the corresponding polyiamond. (D) The boundary points arranged in the order of connectivity of the polyhex. Note that the vertices 30/38 and 31/37 are traversed twice to complete the loop. (E) The boundary points arranged in the order of connectivity of the polyiamond.